\documentclass[preprint,proceedings]{rmaa}

\usepackage{paralist}

\usepackage{psfig,color}

\def\mdot{$\dot{\rm m}$ }

\SetYear{2003}
\SetConfTitle{Compact Binaries in the Galaxy and Beyond}

\title{X-Ray Observations of Classical and recurrent Novae in Outburst} 

\author{
  M. Orio,\altaffilmark{1,2}} 

\altaffiltext{1}{INAF -- Turin Astronomical Observatory, Italy.}
\altaffiltext{2}{Department of Astronomy,  U Wisconsin at Madison, USA.}
\shortauthor{Orio}
\shorttitle{X-ray Observations of Novae}

\fulladdresses{
\item Marina Orio: INAF -- Turin Astronomical Observatory,
              Strada Osservatorio 20, I-10025 Pino Torinese (TO), Italy
         and
Department of Astronomy, 475 N. Charter Str., Madison WI
              53706, USA.
(\email{orio@astro.wisc.edu}) 
}

\listofauthors{M. Orio}
\indexauthor{Orio, M.}
\abstract{I review  X-ray observations of classical and recurrent novae
in outburst, some of them recently done with 
{\sl Chandra} and {\sl XMM-Newton} for 12 objects. 
Significant X-ray flux is emitted by  the nova shell,
 with a peak luminosity up to L$_{\rm x}$=10$^{35}$ erg s$^{-1}$ in
 the 0.2-10 keV range.
In recurrent nova systems, or in novae hosting a red giant, the
source of X-rays may be 
previous circumstellar matter shocked by the
nova wind. However, for most classical
novae, X-rays originate inside
the nebula ejected in the outburst.  The data indicate a very
 high fraction of shocked material, and a non-smooth, varying wind
outflow. A nebular emission line spectrum is also observed at late phases. 
In about half of the observed novae, the central white dwarf appears 
as a very luminous supersoft
X-ray source for 1 to 9 years after the outburst. It is the
 best type of object to study the characteristics of shell hydrogen burning
on white dwarfs in single degenerate systems. 
Still incomplete statistics indicate that  the duration
of the supersoft X-ray phase is peaked around $\simeq$2 years.
The correlation of the X-ray light curve with 
the nova properties is not quite clear.
Recently,  ``template grating spectra'' with high S/N 
have been obtained for  V4743 Sgr.  The X-ray
light curve of this nova reveals a rich and complex power spectrum, 
with signatures of non-radial g-mode oscillations of the
white dwarf.  The oscillations  and the spectra
allow to determine 
 the properties of the shell hydrogen burning white dwarf. 
}

\addkeyword{Stars: binaries, white dwarfs}
\addkeyword{X-rays: stars}

\begin{document}
\maketitle

\section{Mechanisms of X-ray emission from novae}
Classical and recurrent novae are an ancient 
topic in astronomy, traditionally studied by 
optical astronomers. Yet, they emit in
 all wavelengths from gamma to radio, and 
are also an interesting subject of study for the younger science
of X-ray astronomy. Novae may emit X-rays because of 
four different physical mechanisms:
1) Shocks in the ejected wind or between the ejecta and interstellar
 or circumstellar medium, which produce a thermal bremsstrahlung
spectrum, and probably, later, ionize
of the ejecta,  2) Thermal emission of the central white dwarf
(hereafter WD) atmosphere,
 like in supersoft X-ray sources (see Greiner 2000), 3) Resumed accretion
(through a disk or magnetic)
for which a thermal bremsstrahlung spectrum is also observed, 
 (see Orio
 et al. 2001a and references therein) 
4) Finally, X-ray flux may result from Compton
 degradation produced by radioactive decays (Livio et al. 1992).

The third and fourth mechanism are very interesting to
 study the secular evolution of novae and their nucleosynthesis 
 output, respectively. However,  I will review here only
 the first two mechanisms, which are now generally known to occur
in most novae.    
The emission originating from the nova shell 
teaches about the wind emission and the 
the nebular physics, indicating 
electron density, plasma temperatures, and  conditions
of clumping and asymmetry in the ejecta.
The atmospheric emission from the central WD offers
 instead a unique possibility to derive the physical parameters 
of an extremely hot white dwarf which is 
burning hydrogen in a shell. A single-degenerate binary system
 hosting such a WD  is a potential type Ia SN progenitor. 
 It is still a matter of debate whether recurrent novae
are statistically significant as  
type Ia SN progenitors, and we know that only few classical novae
may be. However, post-outburst nova WD are 
of great interest for type Ia supernova studies, because 
we seldom have the possibility to observe the atmosphere of  a WD in
these extreme conditions. In other types of supersoft X-ray
 sources, the central WD seems to be often obscured by a wind
 outflow, or an extended disk corona (e.g. Cal 87 described by Greiner et al.,
 and Orio et al., at this conference). 

Novae that appear as supersoft X-ray
sources for   long enough  that 
their atmosphere is no longer totally or partially hidden by the ejecta, are
therefore very important ``templates'' of  shell hydrogen burning  WD.
 In order to assess if
the WD is increasing in mass after each outburst and
the nova we are observing is a rare type Ia SN candidate, 
we need to verify that the luminous supersoft X-ray source is observed for
 a  long enough time to imply a significant amount of 
 hydrogen  retained  after each outburst, and
that the chemical abundances do not indicate the origin
of the burning material   
 in  the ``eroded'' WD interior, rather than in the accreted matter.
However, some classical novae, and
especially recurrent novae (like perhaps U Sco, see Anupama and Dewangan
 2000, but also Iijima 2002), 
may also become type Ia SN via a sub-Chandra mechanism of explosion
accumulating
a thick helium buffer (see Livio 2003, this conference).  
Since we need to know the abundances in
 the burning layer in both cases,
 high S/N grating observations are essential. 
         
\begin{figure}[!t]
  \includegraphics[width=5.7cm,angle=-90]{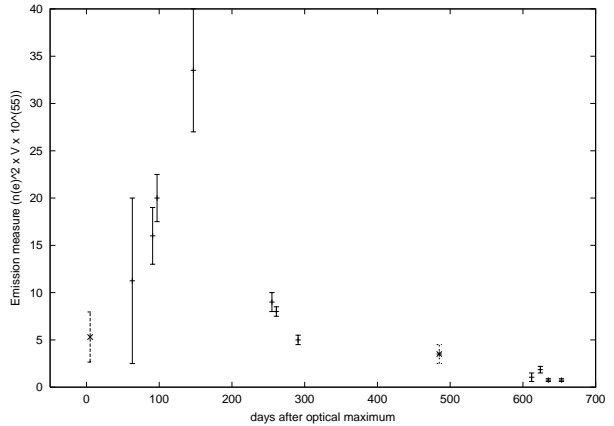}
  \caption{ The emission measure measured with
 {\sl ROSAT} at different post-outburst epochs
for V1974 Cyg (crosses), V838 Her (x), and V351 Pup (asterisk).
}
  \label{fig:simple}
\end{figure}
\section{The early observations}
Novae in outburst were not pointed at with {\sl Einstein}  and
previous X-ray satellites. 
NQ Vul and V1500 Cyg were serendipitously observed and not detected 
(a dubious 1-$\sigma$ detection is recorded for V1500 Cyg). 
\"Ogelman, Beuermann and Krautter (1984, 1987) 
used the {\sl Exosat} LE telescope
to observe three classical novae
in outburst. An initial increase of X-ray flux was
 followed by a plateau and finally a decay. 
 Despite  poor S/N and spectral  
resolution,  the authors suggested  that the X-ray emission was due to
the central WD. The light curve of RS Oph was also monitored 
with {\sl Exosat}, but the X-rays emission was
attributed instead to the nova wind colliding with circumstellar
material, previously emitted in a wind by the red giant
companion (Mason et al. 1987, Contini et al. 1995).

\section{The hot ejected shell}

 Between 1989 and 1999 {\sl ROSAT}
 was used to observe novae  in the 0.2-2.4 keV energy range, with
moderate spectral resolution (with the {\sl PSPC}) and
 5 arcsec spatial resolution (with the {\sl HRI}).
 {\sl ROSAT} 
also contributed many serendipitous observations thanks to the
1 degree field of view of the {\sl PSPC} and to the
 all-sky-survey it performed.
3 out of 7 Galactic and Magellanic Clouds
 novae were observed in the first 2 post-outburst years,
  and showed  X-ray emission of non-atmospheric origin,
``hard'' in the {\sl ROSAT} range. It was attributed
 to thermal  emission in  the ejected shell. 2 more novae seem to have been 
hard sources although they were only observed with the HRI
(see Orio et al. 2001a for a discussion).
2-3 years after the outburst, no correlation was
 found between hard X-ray flux and 
post-outburst time. Probably only accreting sources were still 
observed.   
4-6 years old novae were never detected, suggesting
 shorter cooling times of the shells. The novae observed with
{\sl EXOSAT} were also re-observed. 
GQ Mus was found to be still bright but only supersoft
(\"Ogelman et al. 1993). 
PW Vul had faded, QU Vul was detected but appeared  to have
 become much fainter and it was not supersoft (Orio et al. 2001a).   
Balman et al. (1998) fitted the spectrum of  V1974 Cyg
with a simple double model to disentangle the supersoft
central source and the nebular component. They
concluded that
the X-ray luminosity was L(x)=10$^{33-34}$ erg s$^{-1}$ at
maximum and it decreased in time, like the emission measure in Fig. 1.
The high value of the emission measure
($\approx$n$_{\rm e}^2 \times$V), up to few 10$^{56}$
cm$^{-2}$ at maximum,    indicated a large amount of shocked material.
In Fig. 1 I plot the emission measure as a function of time
including sporadic measurements for two other novae,
V838 Her (Lloyd et al. 1992) and V351 Pup (Orio et al. 1996),
that agree well with the V1494 Cyg curve.

 After V1974 Cyg, a second optically bright nova  became extremely
bright in X-rays, V382 Vel, observed in 1999-2000 
with four different satellites: {\sl Beppo-SAX},
{\sl ASCA}, {\sl Rossi-XTE} and {\sl Chandra}.
This nova was a hard source at 15 days
post-outburst, with a plasma temperature kT$\approx$6 keV (Orio et al. 2001b).
 The emission measure in the 0.2-10 keV
range was a factor 10 larger than observed for V1974 Cyg at maximum in the 
0.2-2.4 keV range.
There was an initial rise in plasma temperature, then a slow
 decay in the next two months. 
 A luminous supersoft component was predominant after 6 months (Orio et 
al. 2002), but not all the supersoft X-ray flux could due to
the central WD. The spectrum could not
 be fitted with any atmospheric model. 
Four months later, only  
emission lines were observed in the supersoft X-ray range 
with the {\sl Chandra-LETG} grating (Burwitz et al. 2002).
Orio et al. (2002) concluded 
that the emission lines, of nebular 
origin, were already present and  superimposed on the atmospheric
continuum of the WD at the peak of the supersoft X-ray phase,
 making it difficult to disentangle the atmospheric
component. 
  Mukai and Ishida (2001) analysed observations done
 in the first two months and measured  a peak luminosity
L(x)$\simeq 10^{35}$ erg cm$^2$. The emission measure in the 1-10 keV
range was an order of magnitude higher than in the {\sl ROSAT}
range for other novae (see Fig. 1).  These authors
 found constraining evidence that, like for V1974 Cyg, 
N(H) was decreasing in time and discussed 
how shocks must have occurred {\it inside} the ejecta.
 They concluded that the characteristics and evolution of the
spectrum  did not indicate  a  post-maximum wind 
colliding with material from pre-maximum wind,
but rather a non-smooth outflow with varying velocity,
 which may certain prove to be a challenge for the models.

  5 new novae observed within 2 years from the outburst
with {\sl Chandra} and/or {\sl XMM-Newton}
 showed  non-atmospheric X-ray emission. For a nova at 
known LMC distance (LMC 2000, Greiner et al. 2003) 
a luminosity L$_{\rm x} \approx$
5 $\times$ 10$^{-34}$ erg cm$^{-2}$ s$^{-1}$ was measured in the 0.2-10 keV 
51 days after the outburst.  This 
is important because Galactic novae distances are often known with
poor precision.
The recurrent novae IM Nor and CI Aql were 
X-ray sources without any indication
of emission from the central star (Greiner \& DiStefano 2002,
Tepedelenlioglu et al.  2004), and
3  out of 5 novae observed with {\sl XMM-Newton} 2 to 4 years
 post-outburst showed only non atmospheric emission.
For V2487 Oph (1998) 
the X-ray flux is attributed to  ongoing accretion (Hernanz \& Sala 2002).
My analysis or archival
observations of V Sgr 4633 (1998) shows that  
the X-ray emission is likely to be originated in a slowly cooling shell. 
I also found that
LZ Mus (1998) was not an X-ray source 3 years after outburst,
that V4444 Sgr (1999) was only  marginally detected, and that 
V 1141 Sco (1997), if the {\sl
Beppo-SAX} detection was real (see Orio et al. 2001a),
must have  cooled down  and does not emit X rays 3 years later.

An unusual obscuration of the absorption lines spectrum
of the WD in nova  V4743 Sgr, the
main X-ray source showing 
 an atmospheric absorption lines spectrum,  
 was recently observed for in V4743 Sgr  and
revealed the emission like spectrum of a less luminous source
(Ness et al. 2003).  Like for the last observations of V382 Vel, 
this second source is most likely of nebular origin 
(either due to shocks or photoionization by the central
source, see Greiner et al. 2003 for a discussion).
This observation poses a doubt as to whether we
can discriminate between wind 
and WD atmosphere with a CCD-type detector,
 when they are superimposed and at a stage of comparable
 luminosity. 
 For V4743 Sgr we were lucky, because the emission lines source
 was observed to be much less bright then
the central source (Ness et al. 2003), but this is not always so 
at any epoch for any nova. The
evolution of the two sources of soft X-rays,
the nebula and the WD, may have a peak in luminosity
around the same time.  Spatially resolving the shell is not
 a realistic goal in most cases,
because the angular diameter  of the shell at
typical nova distance d$\geq$1 kpc is mostly $<$1 arcsec
in the first two years. 
With CCD-like instruments, deriving the WD parameters may be impossible
because 
we do not resolve in the spectrum the nebular emission lines 
 but only observe a composite ``pseudo-continuum''. 

\begin{table}[!t]\centering
  \newcommand{\DS}{\hspace{6\tabcolsep}}
 \setlength{\tabnotewidth}{0.9\textwidth}
 \setlength{\tabcolsep}{1.33\tabcolsep}
 \tablecols{3}
  \caption{Time to reach supersoft X-ray maximum (t$_{\rm soft}$)
and constant bolometric
 luminosity phase length (t$_{\rm bol}$) for classical and recurrent novae}
 \label{tab:SSX}
 \begin{tabular}{lrr}
    \toprule
NOVA & T$_{\rm SOFT}$ &  T$_{\rm BOL}$ \\
\midrule
GQ Mus      &            &    9-10  years \\
V1974 Cyg   & 434-511  days & 2-3  years \\
V382 Vel    & 59$<{\rm t}\leq$184 days  & $<$265  days \\
V1494 Aql   & 180-210  days   & 2.5-3  years           \\
V4743 Sgr   & 120-180 days   &           \\
N LMC 1995  & $\geq$7 months &   6$<{\rm t}<$8 years\\
N in M31    & $\leq$ 11  months &  2-2.5   years \\
N LMC 2000  & $<$ 51   days & $<$51 days \\
U Sco       & $\simeq$ 20  days   &                   \\
IM Nor      & 30$<{\rm t}<$150 days  &                \\
CI Aql      & $<$16 months     &  $<$ 16 months \\
      \bottomrule
  \end{tabular}
\end{table}
\section{The supersoft central X-ray source}
In Table 1 I show a range of time to reach maximum in supersoft
X-rays and the duration of the supersoft X-ray phase 
(which is also the length of the whole constant
bolometric luminosity phase). 
  I have found no clear correlation
of the times in the table with the parameters used to classify novae,
like t(3) (the time for a decay by 3 magnitudes in optical).
 In Greiner et al. (2003) we explain the 
length of the constant bolometric luminosity
phase in terms of the complex interplay
 between the leftover mass (difference
 between the mass necessary  to ignite the nova flash, which is a function
of the WD mass M(WD) and mass transfer rate \mdot, in turn
dependent on the orbital
period, and the ejected mass M(ej)) and
the minimum mass necessary for thermonuclear burning to continue 
(a function of M(WD)). However, M(WD) is normally
unknown and be inferred from a theoretical correlation
with t(3),  but 
other factors also effect t(3). 
M(ej) is only roughly estimated in
the literature, with huge error bars. It is still difficult to predict 
clearly 
how often, and why, novae retain part of the
accreted envelope  or eject instead more material, namely even WD material.
In my opinion, the situation is in still confusing and 
we cannot yet make statistically significantly
 predictions concerning the end product of nova evolution.
   
 The last three lines of Table 1 show in fact the results known
on recurrent novae (RN). U Sco, a very fast RN, was 
observed to be a supersoft X-ray source only 20 days after
maximum, and it was not observed again (Kahabka et al. 1999). The 
optical evolution of
 CI Aql was significantly slower than the one of U Sco,
 and for this reason Greiner \& DiStefano (2002) proposed
to observe it only 16 months after the outburst, 
 expecting an X-ray light curve like V1474 Cyg,
 however  the supersoft
 X-ray source was not detected. IM Nor, 
another RN which shares many physical
characteristics with CI Aql, was observed with
{\sl Chandra} twice, at  much earlier epochs,  1 and 6 months
post-outburst. Given 
t(3) of the same order of magnitude as CI Aql but  a much shorter 
orbital period (which most likely indicates higher \mdot \ and lower
envelope mass necessary to trigger the outburst),
IM Nor was expected to turn into a supersoft X-ray source
within only half a year. Optical observations
showed that the envelope became optically thin (Tepedelenlioglu et al.,
 2004, in preparation). 
   However, the result was negative again. Most models 
foresee that sufficient  hydrogen to burn is left over
on the WD of RN. RN  are thus 
type Ia SN candidates, because the WD is initially massive, and it 
grows towards the Chandrasekhar mass after repeated outbursts.
  However,  the non-detection in supersoft X-rays indicates that
 RN remain a puzzle. Was CI Aql observed too
late and did IM Nor appear as a supersoft X-ray source only 
 for a short time between the two observations? 
 Perhaps  we will have to conclude
 that the WD mass does not grow, but more statistics and more frequent 
sampling of RN light curves are necessary. 

V1974 Cyg, observed with {\sl ROSAT} in 1992-1993, proved that Galactic
  novae can be the brightest supersoft X-ray sources, 
and actually some of the brightest among all  X-ray sources.
Two novae observed with ROSAT  and other satellites as
 supersoft X-ray sources for longer than others: GQ Mus and N LMC 1995.
N LMC 1995 was observed again with
{\sl XMM-Newton} and still observed as a supersoft X-ray source
6 years after the outburst (Orio et al. 2003). Atmospheric models
 did fit the observed X-ray spectrum, apparently
not obscured any more, even partially, by nebular emission.
However, this nova was not sufficiently bright for grating
spectra, so important parameters remained unconstrained. 
 A recent
follow up observation with {\sl XMM-Newton} (Orio et al. 2004,
 in preparation) showed that the supersoft X-ray source however
almost faded after 8 years.

We still have
only small number statistics, but TOO observation done with
 {\sl Chandra} by a large international collaboration 
indicate that probably about half of all 
novae turn into super-bright sources, at least as 
luminous as V1974 Cyg, for a few months.
We do not know whether the supersoft X-ray source phase is
 extremely short lived in other novae or does not occur. 
The light curves of V382 Vel, V1494 Aql and V4743 Sgr  show 
intense variability that is not clearly understood. For V382 Vel,
 there were short obscurations (Orio et al. 2002). 
 In the case of V1494 Aql a flare was observed, which lasted
for about 20 minutes. The count rate increased by a factor of 
10. V4743 Sgr (see also Krautter 2003, this conference) 
a bright nova that reached
V(max)=5 on September 20 2002, was
observed in X-rays in November of 2002 (with {\sl Chandra ACIS-S}),
on  March 20 2003 (with {\sl Chandra LETG+HRC}), in
April 4 2003 (with {\sl XMM-Newton}), and in June and September of 2003
(again with {\sl Chandra LETG+HRC}.) The nova
had flared up by March 2003: at the time
 of the second observation, for 3.6 hours the count rate
was an astonishing 40 cts s$^{-1}$ with the {\sl LETG}, until the
 obscuration mentioned in Section 3 started.
This dramatic fading was no longer detected in April, July
and September, although it April the observation was {\it longer}
 than the optical period, so an eclipse seems to be ruled out.

The most interesting discovery in
the nova X-ray light curves are the non-radial 
oscillations of the WD, first proposed to explain the power
 spectrum of V1494 Aql (Drake et al.
 2002) and then observed, I dare say beyond doubts,
 for V4743 Sgr. 
   In  Leibowitz et al. (2003), we
discuss the power spectrum of this nova. The two highest peaks correspond to
the nearby
periodicities 1308 and 1374 s, respectively. A 1325 s
 period detected in the March {\sl Chandra} light curve
(Ness et al. 2003), may be  
due to the interference of these two periods. While
the 1308 s period has a sinusoidal structure, the 1374 s is
not symmetric. Its first overtone also appears in the power spectrum, but
it is energy dependent and
almost disappears in the light curve is extracted in the energy
range 0.2-0.4 keV (see Fig. 7  in Leibowitz et al. 2003). 
In addition there are many
more periodicities, as it is true for pulsating
PG1059 stars and WD in general. The rich 
power spectrum,  shown in Fig. 4 in Leibowitz et al. (2003),
 appeared to change 
somewhat in the subsequent observations in the summer of 2003. 

Blue-shifted absorption features corresponding
to high ionization states appear in both the
{\sl Chandra} and {\sl XMM-Newton} spectra. While the velocity was 
$\simeq$2400 km/s in March spectrum,
absorption features with lower velocity are
also identified in the April {\sl XMM-Newton} spectrum.
 The blue shift seems to indicate 
 that the nova wind starts at the very base of the WD atmosphere.
In addition to the absorption features,
 some of the emission features due to the  nebula can be
 distinguished as they are superimposed  on the atmospheric 
spectrum. A preliminary analysis of the {\sl XMM-RGS}
 spectra of April of 2003 indicates that
 carbon was not very abundant, that the effective temperature was 
 T$\approx$450000 K, and that effective gravity  
 was g$\leq 10^8$ cm s$^{-2}$. 

The non radial WD pulsations may prove to be an additional
 and important way to derive
    WD parameters. If we succeed in 
 developing detailed models, these
 oscillations, which are easily detectable in the X-ray light curve,
 are a promising possibility to probe T, g and the abundances  of 
 the WD, other than by using WD atmospheric models, thus yielding 
 independent means to derive these parameters.
%



\begin{thebibliography}
\bibitem{} Anupama, G.C., \& Dewangan, G.C. 2000, AJ, 119, 1359 
\bibitem{} Balman, S., Krautter, J., \& \"Ogelman H., 1998, ApJ, 499, 395
\bibitem{} Burwitz V., Starrfield, S., Krautter, J. \& Ness, J-U. 2002, in
Classical Nova Explosions, AIP Conf. Proc. Vol. 637, M. Hernanz \& J. Jose eds.,
p. 377
\bibitem{} Contini M., Orio M., \& Prialnik D. 1995, MNRAS, 275, 195
\bibitem{} Drake, J., et al. 2003, ApJ, 584, 448
\bibitem{} Greiner, J. 2000, New Astronomy, 5/3, 137
\bibitem{} Greiner, J., \& DiStefano, R. 2002, ApJ, 578, L59
\bibitem{} Greiner, J., Orio, M., \& Schartel, N. 2003, A\&A, 405, 703
\bibitem{} Hernanz, M. \& Sala, G. 2002, Science, 298, 393
\bibitem{} Kahabka, P., Hartmann, H.W., Parmar, A.N., \& Neguerela, I.
1999, A\&A, 347, L43 
\bibitem{} Iijima, T. 2002, A\&A, 387, 1013 
\bibitem{} Leibowitz, E., Orio, M., et al. 2003, preprint 
\bibitem{} Livio, M., Mastichiadis, A., Ogelman, H., \& Truran, J.W. 1992,
ApJ, 394, 217 
\bibitem{} Lloyd, H.M., et al. 1992, Nature, 356, L222 
\bibitem{} Mason, K.O., Cordova, F.A., Bode, M.F., \& Barr, P. 1985, in: RS Oph (1985) and the Recurrent Nova Phenomenon, ed. M.F. Bode,
 VNU Science Press (Utrecht), 167
\bibitem{} Mukai, K., \& Ishida, M. 2001, ApJ, 501, 1024 
\bibitem{} Ness J-U., et al. 2003, ApJ, 594, L127
\bibitem{} \"Ogelman, H., Krautter, J. \& Beuermann, K. 1984, 287, L31
\bibitem{} \"Ogelman, H., Krautter, J. \& Beuermann, K. 1987, A\&A, 177, 110
\bibitem{} Orio, M., Covington, J., \& \"Ogelman, H. 2001a, 373, 542 
\bibitem{} Orio, M., Balman, S., Della Valle, M., Gallagher, J.S., \&
 \"Ogelman, H. 1996, ApJ, 463, 221
\bibitem{} Orio, M., et al. 2001b, MNRAS, 326, L13
\bibitem{} Orio, M., et al. 2002, MNRAS, 333, l11 
\bibitem{} Orio, M., Hartmann, W., Still, M., \& Greiner, J. 2003,
 ApJ, 594, 435 
\bibitem{} Tepedelenlioglu, E., Orio, M. Starrfield, S., et al.
 2004, in preparation 
\end{thebibliography}
\end{document}